\expandafter\ifx\csname natexlab\endcsname\relax\fi

\documentclass[12pt,preprint]{aastex}
\usepackage{hyperref}
\def\a4{\hsize 17.0cm \vsize 25.cm}

\shorttitle{}
\begin{document}

\title{The realm of the Galactic globular clusters and the mass of their primordial clouds}

\author{Guillermo Tenorio-Tagle\altaffilmark{1}, Casiana Mu\~noz-Tu\~n\'on\altaffilmark{2},   
Santi Cassisi\altaffilmark{3}, Sergiy Silich\altaffilmark{1} }

\altaffiltext{1}{Instituto Nacional de Astrof\'\i sica \'Optica y
Electr\'onica, AP 51, 72000 Puebla, M\'exico; gtt@inaoep.mx}

\altaffiltext{2}{Instituto de Astrof\'\i sica de Canarias
cmt@iac.es.}

\altaffiltext{3}{INAF – Astronomical Observatory of Collurania, via M. Maggini, 64100 Teramo, Italy; cassisi@oa-teramo.inaf.it 
}

\begin{abstract}

By adopting the empirical constraints related to the estimates of Helium enhancement ($\Delta Y$), present mass ratio between first and second stellar generations ($M_{1G}/M_{2G}$) and the actual mass of Galactic globular clusters ($M_{GC}$), we envisage a possible scenario for the formation of these stellar systems. Our approach allows  for the possible loss of stars through evaporation or tidal interactions and different star formation efficiencies. In our approach  the star formation efficiency of the first generation ($\epsilon_{1G}$) is the central factor that links the stellar generations as it not only defines both the mass in stars of the first generation and the remaining mass available for further star formation, but it also fixes the amount of matter required   to contaminate the second stellar generation. In this way, $\epsilon_{1G}$ is fully defined by the He enhancement between successive generations in a GC.  We also show that globular clusters fit well within a $\Delta Y$ {\it vs} $M_{1G}/M_{2G}$ diagram which indicates three different evolutionary paths. The central one is for clusters that have not loss stars, through tidal interactions, from either of their stellar generations, and thus their present  $M_{GC}$ value  
is identical to the amount of low mass stars ($M_* \le$ 1 M$_\odot$) that resulted from both stellar generations.
Other possible evolutions imply either the loss of first generation stars or the combination of a low star formation efficiency in the second stellar generation and/or a loss of stars from the second generation. From these considerations we derive a lower  limit to the mass ($M_{tot}$) of the individual primordial clouds that gave origin to globular clusters.

  \end{abstract}

\keywords{galaxies: star clusters --- Globular Clusters --- Supernovae
          Physical Data and Processes: hydrodynamics}

\section{Introduction}
\label{sec:1}

During the last decade, one of the most astonishing results in the context of
Stellar Astrophysics has been the discovery of the multiple population 
phenomenon in Galactic Globular Clusters (GGCs), i.e. the presence of distinct
sub-populations characterized by their specific chemical patterns 
\citep[][and references therein]{Piotto2012,Bellini2010,Carretta2009,
Marino2008,Marino2012,Renzini2015}.

The most distinctive chemical features which characterize the various
stellar populations in a given GC are the presence of different abundances of
light elements (such as C,N,O and Na) and different level of He enhancement, 
from extremely small $\Delta{Y}\approx0.01$ \citep{Milone2012B,Piotto2013},
to huge values $\approx0.15$ \citep{Norris2004,Piotto2005,Piotto2007,King2012,
Bellini2013}.
These distinct chemical patterns among the various sub-populations in the same
cluster, create the well-known O-Na anti-correlation that is - from a chemical
point of view - the most evident property of the Multiple Population 
phenomenon \citep[see, e.g.,][]{Gratton2012,Marino2014}. 

At the same time, very accurate photometric investigations performed by means 
of Hubble Space Telescope have revealed the existence of multiple evolutionary
sequences in the Color-Magnitude Diagrams (CMDs) of  various GCs, such as 
distinct Main Sequence (MS), Sub-Giant Branch (SGB), and multiple Red Giant 
Branch (RGB) loci. Actually, the features observed in the CMD change 
significantly from cluster to cluster, and their properties strongly depend on
the adopted photometric systems \citep[see][and references therein]{Milone2010,Milone2012A,Milone2013,Milone2015}. 

The most common - although still debated - scenarios postulate that, in any 
GC, a second (and in some cases also more) generation(s) of stars can form 
from the ejecta of intermediate-mass and/or massive stars
belonging to the first stellar population, whose initial chemical composition is modified by high-temperature
proton captures.  Although some amount of dilution between pristine (unpolluted) matter
and matter processed through thermo-nuclear burning seems to be unavoidable in order to explain  the trend of the chemical patterns observed, for instance, along  the O-Na anti-correlation diagram (see the discussion in 
\citet{DErcole2011} and \citet{Renzini2015}.

The ability to trace both spectroscopically and photometrically the
various sub-populations hosted by each individual GC, allows now for
the identification of both the primordial stellar component
(the first generation, 1G) and the second generation (2G) stars. At the same time, the 
strong correlation that exists between the spectroscopic signatures of the distinct sub-populations
and their distribution along the multiple CMD sequences, clearly indicate that the peculiar
chemical patterns of 2G stars have to affect both the evolutionary properties of these
stars as well as their spectral energy distribution 
\citep{Sbordone2011,Cassisi2013A,Cassisi2013B,Dotter2015}.

 Although, the photometric and spectroscopic properties of the sub-populations in
GGC sample are nowadays well known, we still face a {\sl pletora} of issues related to
many aspects of the Multiple Population phenomenon. The most important ones being: {\sl i)} the nature of
the polluters, i.e. the type of stars responsible for the nucleosynthesis at the basis of the
2G chemical patterns; {\sl ii)} How 2G stars were actually able to form in the inner regions of GGCs i.e. in a region with a very high stellar density; {\sl iii)} the current fraction of 2G to 1G stars observed in individual
cluster raises the mass budget problem, i.e. since the candidate polluting stars make up only 
a very small fraction of the total  initial mass of the GC, currently observed 1G stars
(and their already evolved companions) are not enough to have produced the required amount
(also accounting for dilution) of enriched matter for a 2G. For a detailed review on the various
problems related to the interpretation of the multiple population phenomenon in GGCs we refer to the work by \citet{Renzini2015, Bastian2015A, Bastian2015B}
and references therein.
 Due to the existence of all these issues related to the explanation of this phenomenon
it is mandatory to fully exploit the empirical constraints derived from the on-going
spectroscopic and photometric surveys \citep[e.g.][]{Piotto2015}
in order to obtain the most reliable possible set of
constraints on the mass of the cluster progenitor, the star formation efficiency for the
various sub-population as well as an estimate of the capability of the various clusters to
retain stars belonging to the distinct sub-populations.

Here we use current empirical estimates on the actual cluster mass, the relative
fraction of 1G and 2G stars as well as current estimates of the He enhancement between
the various sub-populations in a sample of GGCs in order to provide some clues on the
 total mass of the primordial cloud, and at the same time on the possible evolutionary paths
- in terms of the star formation efficiency and the capability of  clusters to retain the stars
of the various sub-groups - followed by the clusters in our sample.

\section{Secondary stellar generations in Galactic globular clusters}

Let us assume a massive cloud with a total mass $M_{tot}$ that collapses and 
forms a first stellar generation with an efficiency $\epsilon_{1G}$. 
We also assume that the event leads to a full \citet{Kroupa2001}
initial mass function (IMF) with stars in the range 0.1 - 120 M$_\odot$ and, 
as expected from massive  starbursts restricted to a small volume, a large 
fraction of their massive stars should end up as interacting binaries 
\citep[see][]{DeMink2009, Schneider2014}.
The major implications of this \citep[as in][]{DeMink2009, Izzard2013}
is that instead of massive stars producing powerful winds during all their
life most of their H burning products would exit the stars with a low velocity
($\sim$ a few tens of km s$^{-1}$) and thus, during the early evolution, the 
large collection of massive binaries are likely to hold the remaining cloud 
against gravitational collapse while contaminating the gas left over from star
formation without disrupting its centrally concentrated density distribution. 
This latter point 
is also an important issue as, under such conditions, the blast waves from sequential supernovae (SNe) are all likely to undergo blowout. i.e. the sudden acceleration of the blast wave and of its shell of swept up matter as soon as they enter the steep density gradient present in the remaining cloud. Such events lead to Rayleigh-Taylor instabilities in the swept up shells and to their fragmentation, favouring the venting of the SN debris out of the remaining cloud without causing its contamination\footnote{It is worth noting that a different scenario has been recently envisaged by \citet{Calura2015}
on the basis of 3D hydrodynamic simulations. In this scenario, the integrated feedback from stellar winds and SNe explosions would be responsible for the pristine gas removal within $\sim14$~Myr since the cluster formation.} 
\citep{TenorioTagle2015,Krause2012}.
Only the most massive clouds with a mass $M_{gas} \ge 10^6$ M$_\odot$
would be able to retain the high velocity ejecta of a few percent of the supernova explosions from first generation and only if these explode near the center of the gas distribution. In these latter cases, the mass available for a second stellar generation would also be contaminated with supernova products and its resultant stars will be likely to show a prominent Fe spread instead of the  uniform metallicity characteristic to less massive clusters 
\citep[see the discussion in][and references therein]{Marino2011,Milone2015}.
%Marino A. F. et al., 2011, and Milone et al. 2015 and references therein). 

All these conditions here assumed, seem to be at work in young, massive compact clusters such as NGC 5253, II Zw 40, Henize 2-10, all completely buried in their natal clouds, which only allow for their detection at infrared and milimeter wavelengths \citep[see][and references therein]{Turner2015,Beck2015}.
These assumptions seem a resonable possibility to justify secondary stellar 
generations in the young compact clusters such as NGC 1569-A, NGC 1569-B, 
NGC 1705-1, etc. \citep[see][]{Larsen2011} 
for which their resolved HST-ACS photometry implies secondary stellar generations separated from their first generation by an age of less than 50 Myr.  

Following gravitational collapse, the mass in stars of the first  stellar 
generation is:

\begin{equation}
\label{eq1}
M_{1G} = \epsilon_{1G} M_{tot} 
\end{equation}

\noindent If one accounts for tidal interactions and the possible loss of stars during the lifetime of a GC, then one can multiply equation 1 by  $\alpha_{1G}$ ($\le$ 1) to account for the fraction of first generation  stars that are never lost from the cluster and contribute to the present mass.

%--------------------------------------------------------------------------
\begin{figure}[htbp]
\plotone{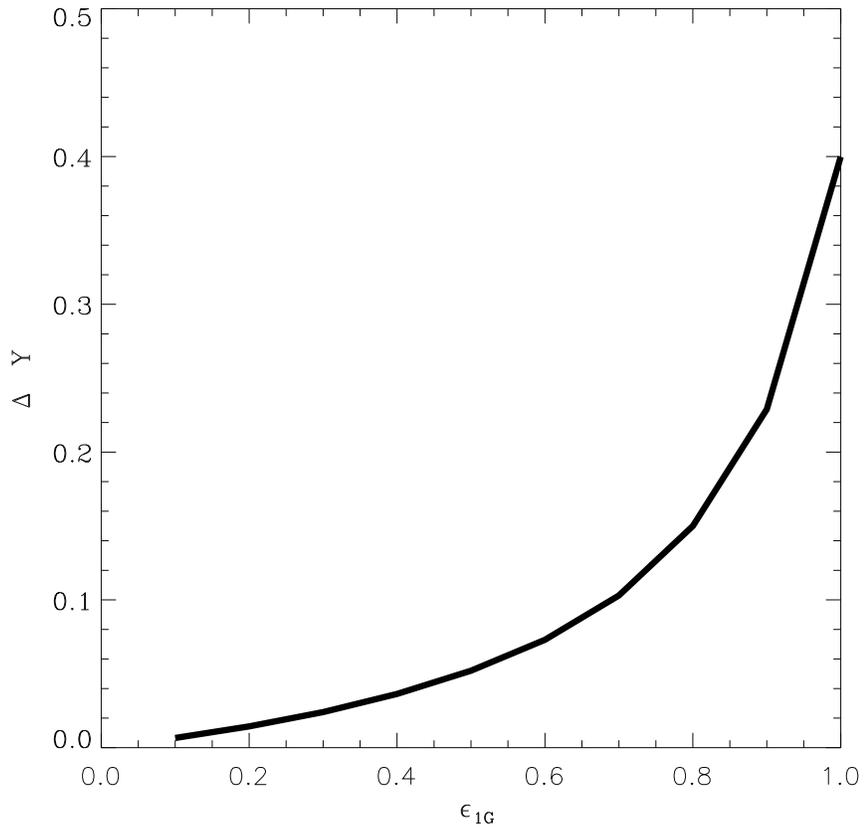}
\caption{$\Delta{Y}$ {\it vs} the star formation efficiency of the first stellar generation ($\epsilon_{1G}$).}
\label{fig:1}
\end{figure}
%------------------------------------------------------------------------------

We shall further assume that the left over gas becomes thorougly contaminated by the H burning products shed by stars from the first stellar generation. 
The resultant He enhancement  $Y$  in the left over cloud is then:

\begin{equation}
\label{eq4}
Y = \frac {M_pY_p + M_cY_c}  {(M_p + M_c)} 
\end{equation}

\noindent where the subscripts {\sl p} and {\sl c} stand for the pristine and the contaminer gas, respectively. Following \citet{DeMink2009}
the total mass of the contaminer amounts to $\sim 0.15 M_{1G}$ which accounts for the fraction of the stellar mass shed by interacting massive binaries from the 1G. Thus the contaminer mass is 
$M_c = 0.15 M_{1G} = 0.15 \epsilon_{1G} M_{tot}$
and the mass in pristine gas after the 1G has formed is:
$M_p = (1 - \epsilon_{1G}) M_{tot}$.

In this way  the resultant value of $Y$ is:
\begin{equation}
\label{eq5}
Y = \frac {(1 - \epsilon_{1G}) Y_p  + 0.15 \epsilon_{1G} Y_c}    {(1 - \epsilon_{1G}) + 0.15 \epsilon_{1G}}
\end{equation}

\noindent and the corresponding value of $\Delta Y$:

\begin{equation}
\label{eq4}
\Delta Y = Y - Y_p = \frac {0.15 \epsilon_{1G} (Y_c - Y_p)} {(1- \epsilon_{1G} + 0.15 \epsilon_{1G})}
\end{equation}

\noindent This is plotted in Figure 1 as a function of $\epsilon_{1G}$, 
assuming as in \citet{DeMink2009} and in \citet{Bastian2013}, 
that $Y_c$ = 0.64\footnote{This value corresponds to the extreme He abundance in the ejecta of their model. However, as we discuss in the following, a change of this value could be easily accounted in the analysis performed in this
work.} and the primordial He abundance  $Y_p$ is equal to 0.24.
Within this framework a strong dilution of the contaminer mass happens naturally for cases with small values of $\epsilon_{1G}$, leading to small values of $\Delta Y$ and conversely, very little dilution is achieved in cases with a large $\epsilon_{1G}$, causing large values of $\Delta Y$. We note that this expected trend leads to a range of $\Delta Y$ values in good agreement with the range inferred from  observations of GGCs as we discuss in the following.

Within the framework here described, the efficiency of star formation of the first stellar generation ($\epsilon_{1G}$) is what defines both the total amount of pristine gas left over from the formation of a first stellar generation as well as the mass of the contaminer gas  ($0.15 M_{1G}$) to lead, upon a thorough mixing, to a contaminated cloud ready to trigger a second stellar generation with its own He abundance $Y_{2G}= Y_p + \Delta Y$ (see equations 3 and 4 and Figure 1). From equations 2 - 4 and the values of $Y_p$ and $Y_c$ there given, $\epsilon_{1G}$ is:

\begin{equation}
\label{eq7}
\epsilon_{1G} = \frac{\Delta Y}{0.06 +0.85 \Delta Y}
\end{equation}

\noindent and thus it is fully defined by the observations. For those 
clusters hosting distinct sub-populations, each with an specific value of 
Y, equation \ref{eq7} should be replaced by: $\epsilon_{(j-1)G} = 
\Delta Y_{j,j-1} / (0.15(0.64 - Y_{j-1}) + 0.85 \Delta Y_{j,j-1})$.
If the contaminer $Y_c$ is 0.64 (as we used for massive binaries) and 
$\Delta Y$ is measured between generation j and j-1.
In this way the $\Delta Y$ values between subsequent generations would lead to the efficiency of star formation of all but the last stellar generation. 

In the present analysis we rely on the prescription for $Y_c$ and the fraction of the total 1G mass available as contaminant gas, as provided by 
\citet{DeMink2009}
in their scenario of interacting, massive binary systems. However, we note that this scenario is based on only one model - so additional computations are strongly needed, and {\sl also} that this scenario, as well as other candidate polluters suggested so far - has its own shortcomings as discussed by 
\citet{Bastian2015A} and \citet{Renzini2015}.
 Notwithstanding, in the present analysis we perform an explorative analysis using these model predictions at face value. In principle,  one could select a different polluter but only if they do not disrupt the cloud left over
from star formation. Their implementation requires only to replace  in  our  
equations the corresponding  values of $M_c$ and $Y_c$. 
The same consideration could be applied to the case of AGB stars 
\citep{DAntona2002},
but not if the model assumes  the loss of the matter left over from the 
formation of a first stellar generation (1 - $\epsilon_{1G}) M_{tot}$
as in  \citet{DErcole2008}, 
as  such an assumption  invalidates our equations 3 - 5. Note that, in that scenario of AGB stars as candidate polluters, the amount of pristine gas that has to be accreted in order to dilute the AGB stellar ejecta is a free parameter, whereas in the present analysis, the diluter gas is simply the residual gas from 1G formation. 
%--------------------------------------------------------------------------
%\begin{figure}[htbp]
%\plotone{f2.eps}
%\caption{The $M_{1G}/M_{2G}$ ratio {\it vs} the star formation efficiency of the first stellar generation ($\epsilon_{1G}$).
%This results from equation 6 using $\alpha_{1G}, \alpha_{2G}$ and $\epsilon_{2G}$, all equal to 1.}
%\label{fig:2}
%\end{figure}
%---------------------------------------------------------------------------

In our approach the total mass available for a second stellar generation (2G) 
is:

\begin{equation}
\label{eq1}
M_{residual gas} = (1 - \epsilon_{1G}) M_{tot} + 0.15 M_{1G}
\end{equation}

\noindent where the second term accounts for the mass that contaminates the gas left over from the formation of the 1G. Having all this mass already in place there is no need to invoke, as in other scenarios for GC formation for the accretion of further pristine gas to dilute the cloud that would give origin to a second stellar generation.
 
As in equation 1, this equation has to be multiplied by   
 a star formation efficiency factor  ($\epsilon_{2G}$) to infer the mass of the second stellar generation. Also, if one is to consider the possible loss of stars through tidal interactions, equation 6 should also be multiplied by the fraction of stars from second generation that remain gravitationally trapped within the cluster ($\alpha_{2G} \le$ 1) and contribute to its present mass 
$M_{2G}$. However, $\alpha_{2G}$ should have a value close to 1, given the empirical evidence showing the 2G being commonly centrally concentrated 
 with respect the 1G 
\citep[see the analysis performed by][]{Bellini2009,Lardo2011,Milone2012A, 
Milone2012C,Milone2013,Milone2015}\footnote{but see also 
\citet{Dalessandro2014} for the case of lack of radial gradients in the 
stellar populations of NGC~6362}.

After some algebra equations 1, 4 and 6 yield:

\begin{equation}
\label{eq2}
X = \frac {M_{1G}} {M_{2G}} = \frac{\epsilon_{1G} \alpha_{1G}}  {\epsilon_{2G} \alpha_{2G}  (1-\epsilon_{1G} + 0.15 \epsilon_{1G})}
= \frac {\Delta Y \gamma} {0.15(Y_c - Y_p)}
\end{equation}

\noindent where 

\begin{equation}
\label{eq1}
\gamma = \frac {\alpha_{1G}} {\alpha_{2G} \epsilon_{2G}}  =  \frac {0.15 (Y_c-Y_p) X}   {\Delta Y} = \frac {0.06 X} {\Delta Y} 
\end{equation}

\noindent and thus $\gamma$ is also constrained by the observations.
From the  above equation, $\Delta Y = 0.06 X / \gamma$. This
is plotted on Fig.~2 
against the mass ratio $M_{1G} / M_{2G}$, where 
the solid line depicts the case when $\gamma$ = 1. 
All clusters located to the right of the solid line have a $\gamma$ larger than one, wheras those to the left have a $\gamma$ smaller than one. For example, 
the dotted lines in Figure 2 correspond to values of $\gamma$ = 1.4 and 3.3, whereas the dashed lines correspond to $\gamma$ = 0.5 and 0.7.

The three variables ($\alpha_{1G}$, $\alpha_{2G}$ and $\epsilon_{2G}$) cause a degeneracy in equation 7 and 8.
Nevertheless, the position of a stellar cluster on the $\Delta Y - M_{1G} / M_{2G}$ parameter space allows one to draw some conclusions regarding its formation and evolution. For example, all clusters to the right of the solid line have the product $\alpha_{2G}$  $\epsilon_{2G}$ smaller than $\alpha_{1G}$. This implies that in all such clusters the second generation was formed with a small efficiency: $\epsilon_{2G} <  \alpha_{1G} < 1$, as the value of $\alpha_{2G}$ 
should be close to 1. On the other hand, all clusters to the left of the solid line in Figure 2 have  $ \alpha_{1G} < \epsilon_{2G} < 1$ (as $\alpha_{2G} \approx 1$) what implies that during the evolution such clusters lost
a significant fraction of their first generation stars.
 
%--------------------------------------------------------------------------
\begin{figure}[htbp]
\plotone{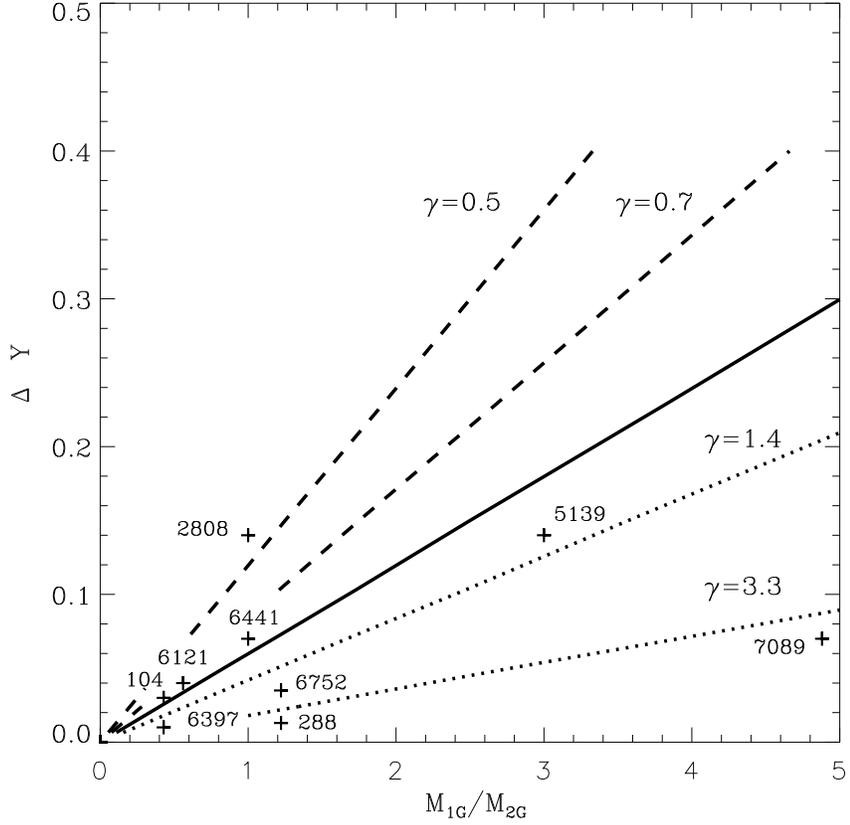}
\caption{The realm of the GGCs. $\Delta Y {\it vs} M_{1G} /M_{2G}$ for  a sample of Galactic globular clusters. The
solid line presents the results from equations 4 and 7 assuming $\alpha_{1G}, \alpha_{2G}$ and $\epsilon_{2G} $, all equal to 1.
Dashed lines to the left of the solid line are also from equation 7 but for different values of $\gamma$ (= 0.7 and 0.5).
Dotted lines to the right of the solid line are from equation 7 but with the product $\alpha_{2G} \epsilon_{2G}$ equal to 0.7 and 0.3 which lead to $\gamma$ values equal to 1.4 and 3.3. The symbols  represent the data for a sample of GGCs with their NGC identification number.}
\label{fig:2}
\end{figure}
%---------------------------------------------------------------------------

In Fig.~2 we also show empirical data for a sample of Galactic GCs 
\citep[][and references therein]{Milone2012A,Milone2014,Milone2015}.
From  these studies one knows three different variables for each cluster:  the value of $\Delta Y$, the total present mass of the clusters ($M_{GC}$) and the ratio of the number of stars from a 2G to the total number: $N_2/N_{tot}$ = $A$. From the latter we know that $N_2 = A N_{tot}$ and if $N_{tot} = N_1 + N_2$, then
$N_1/N_2  = (1-A) / A$. Given  the extended evolution of GCs, we know one is  dealing today with stars within the same mass range ($\le $1 M$_\odot$) and thus such a ratio is equivalent to the mass ratio $ X = M_{1G} / M_{2G}$, as plotted in Fig.~ 2. Before discussing what kind of information we can extract by comparing the empirical dataset with our model, we wish to comment about the reliability of the adopted, empirical estimates on the 2G to 1G population ratio. When comparing the estimates,  currently available in the literature, about $N_2/N_1$ for a given cluster, one can easily find significant differences. For instance, \citet{Carretta2009}
on the basis of the  Na-O anti-correlation in 15 GCs, conclude that the mass of 1G in GCs amounts to about 30$\%$ of the total mass, what they call intermediate population amounts to almost 60$\%$ and in some clusters  there is an extra $\sim$ 10$\%$ in extreme population \citep[a similar conclusion has been obtained by][]{Bastian2015B}.
Seven of their GCs coincide with our sample. However, data in our 
Fig.~2 (and  in Table~1), five of the GCs in our sample have a 1G mass 
larger than the mass of the 2G. Two have an
equal mass and only three of them have a 2G more massive than the 1G. It is also true that two  (NGC 7089 and NGC 5139) of our clusters with a 1G more populous than the 2G  are not included in the original sample by \citet{Carretta2009}.

The estimates we are using in this work are taken from photometric surveys that
consider a quite larger number of stars with respect the spectroscopic surveys. In addition, usually the spectroscopy surveys
are limited to GC outskirts,  whereas HST based photometric surveys sample (also) the inner core of the GCs, so the presence of radial gradient 
could also help in explaining the difference in the estimates of the $N_2/N_1$ ratio\footnote{Indeed, we expect that the ongoing photometric survey of multiple
stellar population in GGCs based on the use of ultraviolet photometric filters 
\citep[which represent a formidable tool for tracing the distinct sub-populations; see][]{Piotto2015}
for a detailed discussion on this issue) can provide more robust empirical estimates on both $\Delta{Y}$ and the 1G/2G population ratio.}.

In Table~1 we list the observed and derived parameters for nine GGCs. Column 1 gives their NGC identification number, and columns 2-4 the He enhancement $\Delta Y$, the derived $M_{1G}/M_{2G}$ as obtained from the observed values $N_2/N_{tot}$, and their present mass in solar masses ($M_{GC}$). Column 5 gives the values of $\epsilon_{1G}$ as derived from equation 5, while columns 6-9 report the present mass $M_{1G}$ and $M_{2G}$, the predicted values of $\gamma$ (from equation 8), and the lower limit value for $M_{tot}$, respectively.

\section{The mass of the primordial clouds}

From the present values of $M_{1G} / M_{2G} $ and the estimated  mass of a GC, one can infer the present mass of both, first and second generation, i.e.:

\begin{equation}
\label{eq1}
M_{2G} = \frac {M_{GC}} {X+1} \quad \quad  {\rm and}  \quad \quad  M_{1G} = M_{GC} - M_{2G} 
\end{equation}

\noindent these are listed for our cluster sample in columns 6 and 7 of Table 1. Such values are relevant if one wants to calculate the total mass ($M_{tot}$) of the pristine cloud that has given birth to both generations, taking into consideration the possible loss of stars through tidal interaction. In the case of the original first stellar generation, this has to be equal to the present mass $M_{1G}$ plus all the stars loss through tidal interactions: $(1 - \alpha_{1G})\epsilon_{1G}M_{tot}$. This through equation 1 leads to:

\begin{equation}
\label{eq1}
M_{tot} = \frac {M_{1G}}  {\epsilon_{1G} \alpha_{1G}}
\end{equation}

Similarly, one can consider the mass available for a second stellar generation (equation 6) and its likely efficiency of star formation ($\epsilon_{2G}$) and evolution ($\alpha_{2G}$) parameters, which lead to:

\begin{equation}
\label{eq2}
M_{tot} = \frac {M_{2G}}  {(1-0.85 \epsilon_{1G}) \alpha_{2G} \epsilon_{2G}}  = \frac {M_{2G}}  {(1-0.85 \epsilon_{1G}) \alpha_{1G}}  \gamma = \frac {M_{2G} X} {\epsilon_{1G} \alpha_{1G}} 
\end{equation}

Equations 10 and 11, allow for the use of  the present mass $M_{1G}$ or 
$M_{2G}$ to derive the  mass $M_{tot}$ of their primordial clouds. 
To infer $M_{tot}$ in the case of a GC with multiple sub-populations 
becomes more complicated as one requires of diagrams linking $M_{j-1}/M_j$
with the corresponding increment on $\Delta Y$ (similar to Figure 1) for each 
of the sub-populations. We leave a full discussion of such an issue to a 
forthcoming contribution.

Given the degeneracy in equation 7, and the lack of other observable(s), the furthest we can reach in our treatment
is to infer a   lower limit to $M_{tot}$. For clusters to the left of the solid line in Figure 2 we assume the upper limit value for the product  $\epsilon_{2G}$ 
$ \alpha_{2G}$ = 1, what leads through equations 10 and 11  to a lower limit on $M_{tot}$. 
In this case,  $\alpha_{1G}$  is equal to  $\gamma$ for all   $\gamma \le 1$.

On the other hand, if instead one assumes the other extreme case:  that during the evolution no stars are loss from the first stellar generation ($\alpha_{1G}$ = 1),  then  the product  $\epsilon_{2G} \alpha_{2G}$,  implicit in equation 11,  has to present values smaller than $\alpha_{1G}$ to lead to  values of $\gamma = \alpha_{1G} / (\epsilon_{2G} \alpha_{2G}) \ge$ 1 in equation 7. The 
result is then to displace the mass ratio $M_{1G}/M_{2G}$ from the solid line 
in Figure 2 towards the right until the present values of $X$ is reached. 
The derived values of $\gamma$ for all clusters are given in column 8 of Table 1 and the predicted values of the lower mass limit for $M_{tot}$ (through equations 9-11) are given in the last column.

%---------------------------------------------------------------------------
\begin{table}[htp]
\caption{\label{tab:1} Observed and derived parameters for a sample of GGCs}
\begin{tabular}{c c c c c c c c c}
\hline\hline
NGC & $\Delta Y$ & $N_1/N_2$ & $M_{GC}$ &   $\epsilon_{1G}$ & $M_{1G}$ & 
$M_{2G}$ &  $\gamma$ & $M_{tot}$  \\
       &   & $ X \sim M_{1G}/M_{2G}$ &  $10^6$M$_\odot$ &    & (10$^{5}$M$_\odot$)
       &  (10$^{5}$M$_\odot$) & & (10$^{6}$M$_\odot$) \\
   (1)    & (2)  & (3) &  (4) & (5)   & (6)
       &  (7) & (8) & (9) \\\hline
104     &   0.03  &  $ 0.43 $ &   $1.26$ & 0.35& 
           $3.78$ & 8.81 &  0.86  & 1.26 \\
288      &   0.013   &  1.22       & $0.08$ & 0.18 & 
           $0.44$ & 0.36 & $ 5.64$ &  0.24\\
2808      &   0.14   &  $ 1$       & 1.58 & 0.78 &
           $7.93$ & 7.92& $ 0.43 $ &  2.35\\
5139      &   0.14   &  $ 3$       &  3.98& 0.78 &
           $29.86$ & 9.95 & $       1.28 $ & 3.83  \\
6121      &   0.04   &  $ 0.56$   &   $0.06$ & 0.43 & 
           $0.23 $ & 0.40 & $ 0.84 $ & 0.06 \\
6397      &   0.01   &  0.43       & $0.25$ & 0.15 & 
           $0.75$ & 1.76 & $ 2.57$ & 0.52 \\
6441      &   0.07   &  $ 1$   &  1.58  & $0.59$ &
           $7.93$ & 7.92 & $ 0.85 $ &  1.59\\
6752      &   0.035   &  $ 1.22$       & $0.16 $ & $0.39$ &
           $0.87$ & 0.71 & $   2.09 $ &  0.22\\
7089      &   0.07   &  $ 4.88$       & $1.0 $ & $0.59 $ & 
           $8.3 $ & 1.70 & $ 4.18 $ & 1.42 \\

\hline\hline
\end{tabular}
\end{table}
%--------------------------------------------------------------------------

\section{Results and Discussion}

There are several issues regarding the observational data.  Some of the GCs, regardless of their total mass, present a mass $M_{2G}$ larger that  $M_{1G}$ but the opposite, an $M_{1G}$  larger than $M_{2G}$, is also possible. Both trends happen also for clusters with a similar value of $\Delta Y$. Note also clusters with a similar mass ratio but a different $\Delta Y$ value.
In our approach, some of the GCs can only be explained if $\gamma$ acquires large values, implying that  the product $\epsilon_{2G}\alpha_{2G}$ should lead to a  low value, reaching  $\sim$ 0.1 to encompass our selected clusters. On the other hand, some GCs require to expulse through tidal effects a large fraction of their stars from 1G so that  $\gamma$, reaches values $ \sim$ 0.5 to encompass our clusters. 

Taking the observational data at face value, note that only four clusters are to the left of  the solid line in Figure 2 and they all present a mass ratio $M_{1G}/M_{2G} \le $1.  In principle there is no physical restriction for this upper limit ($M_{1G}/M_{2G} \le $1) and thus perhaps more data points are needed.  Three of the four clusters to the left of the solid line (NGC 104, NGC 6121 and NGC 6441) are in fact almost at the solid line (with values of $\gamma  \sim$ 0.85), and have their predicted lower limit mass $M_{tot}$ almost identical to the present total mass ($M_{GC}$). This happens despite the fact that they present a different total mass ($M_{GC}$),  a different mass ratio ($X$) and a different $\epsilon_{1G}$. This indeed supports the interpretation that there has hardly being any evolution or loss of stars through tidal interactions from either generation and that the efficiency of star formation of the second stellar generation was rather large (see equation 7). 
The case of NGC 2808 is different as it implies a value of $\gamma \sim$ 0.43) and thus despite its value of $X$ = 1, it most have suffered a major loss of stars from its first generation, which agrees with having a predicted lower limit total mass ($M_{tot}$)  larger than the present mass of the cluster. 

The other five clusters in the sample  are to the right of the solid line and cover a large range of $X$ values. One can compare 
NGC 2808 with  NGC 5139. Both have the same value of $\Delta Y$ and thus the same $\epsilon_{1G}$. However, the former one is to the left of the solid line and the latter to the right, with a mass ratio $X$ = 3. The latter is the only cluster in the whole sample for which the predicted $M_{tot}$ is slightly smaller (by about $4\%$) than the total present mass of the cluster. However, having  $M_{1G}$ larger than $ M_{2G}$ the implication is that most likely no stars were lost  from either generation and its low value of $\alpha_{2G} \epsilon_{2G}$ comes from a low value of $\epsilon_{2G}$ ($\sim$ 0.77) which leads 
to a $\gamma$ = 1.29. 
A similar case is that of NGC 6441 and NGC 7089, both with the same value of $\Delta Y$. 
As mentioned above, NGC 6441 has hardly evolved at all and thus its predicted $M_{tot}$ is almost identical to the observed mass $M_{GC}$. The location of NGC 7089 with an $M_{1G}$ much larger than $M_{2G}$ implies a very low efficiency in the second stellar generation perhaps due to the dispersal of a large amount of contaminated gas (several times $10^5$ M$_\odot$) which were unable to enhance the mass of the 2G. 

To appreciate the range of possible evolutions one should consider also NGC 6441 and NGC 7089 both presenting  a  $\Delta Y = 0.070$. They both have a similar predicted lower limit mass $M_{tot}$ and both gave origin to very similar massive first generations. However, NGC 6441 led to a massive second stellar generation (and present a mass ratio $X \sim$ 1), while in NGC 7089 the second generation is almost five times smaller than the first. 

Low mass clusters such as NGC 6397 and NGC 288  present the lowest values of $\Delta Y$ and thus the lowest  $\epsilon_{1G}$ = 0.146 (see Table 1). If one deducts from the predicted inoitial total mass the mass of the observed first generation, one would have, in principle, an estimate of the mass left for a second stellar generation. NGC 288, with a left over mass $\sim 1.99 \times 10^5$ M$_\odot$ only used less than 20$\%$ of it for its 2G, reaching an $X$ value of 1.22, while NGC 6397 with a mass available 
$\sim 4.4 \times 10^5$ M$_\odot$ transformed almost 40$\%$ of it on its 2G and has an $X$ value equal to 0.43.

In summary: the location of GGCs in the $\Delta Y$ {\it vs} $M_{1G} / M_{2G}$ points in the cases close to the solid line (see Figure 2) to a lack of evolution. Their predicted lower mass limit ($M_{tot}$) is almost identical to the present mass of the GCs ($M_{GC}$).
We have also pointed at clusters to the left of the solid line which can only be justified if there is a major loss of stars from the first stellar generation. And finally, all clusters to the right of the solid line require, most likely, of a low star formation efficiency for the second stellar generation ($\epsilon_{2G}$) to justify the range of values of their mass ratio. All these possibilities may have different origins. GCs have clearly being affected throughout their evolution in different manners, perhaps due to some initial structural parameters, such as the size of the star formation event, and/or by the location of their orbit around the Galaxy centre, that may have allowed for  very different histories of tidal interactions.

Note that so far we have only considered stars with a mass $M_*  \le 1$ M$_\odot$, the stars we see today. However, to  account for the full IMF required to cause the contamination of the 2G and also its full IMF,  
one would have to add an extra 62$\%$ to
the derived values of $M_{tot}$ (see Table 1) to obtain the 100$\%$ original mass of the clouds that gave birth to
the original massive starbursts that after their extended evolution appear today as Galactic Globular Clusters.  The range is then defined in our sample by NGC 6121 and NGC 5139 which indicate the full mass of the primordial clouds ranging from 1.66 $\times 10^5$ M$_\odot$ to 1.01  $\times 10^7$ M$_\odot$.

\section{Concluding remarks}

We depart from a massive cloud which after gravitational collapse gives origin to a first stellar generation with a full IMF.
Some of these stars shed their H burning products and contaminate the gas left over from star formation. We have used for this, the products from massive interacting binaries, assuming that about 15 $\%$ of the total mass of the IMF becomes then available to contaminate the remaining cloud. Other contaminers may also be used, changing in our equations the 
amount of mass ($M_c$) and the appropriate $Y_c$ value expected from these other alternatives. The remaining cloud is assumed to be strongly centrally concentrated with a gas distribution that promotes the blowout of supernova blast waves from first generation massive stars and thus the exit of the ejected matter out of the cluster volume \citep[as in][]{TenorioTagle2015}.

In our models the efficiency of star formation of the first stellar generation ($\epsilon_{1G}$) is the variable that fully defines and links the stellar generations in a GC. First because it defines the amount of stars in the 1G. Secondly, because it fixes concurrently the amount of gas left over while defining the mass available for its contamination. Furthermore, we have shown that $\epsilon_{1G}$  is fully defined by the observations as it only depends on $\Delta Y$ (see equation 5).

We have also shown that the location of GCs in a $\Delta Y$ {\it vs} $M_{1G}/M_{2G}$ diagram provides information 
about their formation and evolution. Information determined by $\gamma$ (in equations 7 and 8), a variable fully defined by the observations (see equation 8). Clusters to the left of the solid line in Figure 2, can only be there if the cluster looses, through tidal interactions, a good fraction of its 1G stars. On the other hand, clusters to the right of the solid line may only lie there as a result of a poor efficiency of stellar formation on their second stellar generation. Fact that is re-inforced by the empirical evidence of having the second generation strongly centrally concentrated and thus clearly unable to have lost its stars.  Clusters at or near the solid line on Figure 2 do not loose stars from either generation as confirmed by the fact that their present mass is equal to the here predicted lower limit total mass  of their primordial cloud ($M_{tot}$). This accounts only for the low mass stars ($ 38\%$ of the full IMF) and thus to obtain a true $M_{tot}$ one should add an extra 62$\%$ to have the full mass of the primordial clouds that gave origin to GGCs. 

On the basis of the present analysis, we suggest the use of the $\Delta Y$ versus 
$M_{1G}/M_{2G}$ diagram as a powerful tool for tracing the formation and 
evolution properties of Galactic GCs.

\section{Acknowledgments}

This study was supported 
by CONACYT - M\'exico, grants 167169, and 131913  and by the Spanish Ministry 
of Science and Innovation  for the ESTALLIDOS collaboration  
(grants AYA2010-21887-C04-04 estallidos4 and 
AYA2013-47742-C4-2-P estallidos5). GTT and SC acknowledge the C\'atedra Severo 
Ochoa at the Instituto de Astrof\'isica de Canarias and GTT also thanks the Jesús Serra Foundation and  
the CONACYT grant 232876 for a Sabbatical leave. SC acknowledges  financial 
support from PRIN-INAF 2014 (PI: S. Cassisi), the friendly hospitality at the IAC, and very helpful discussion with M. Salaris. The authors appreciate 
the science and discussions among the participants of  the ESTALLIDOS
Star Formation Feedback Workshop (IAC, Nov. 2014) which triggered a good 
number of ideas.

\bibliographystyle{apj}
\bibliography{GC}

\end{document}